\documentclass[12pt,dvips]{article}
\usepackage{amssymb}
\usepackage{epsfig,epic,eepic}

\def\u#1{{\underline{#1}}}
\def\part#1{{\partial\over\partial #1}}
\def\maplongright#1{\smash{\mathop{\hbox to .45in {\rightarrowfill}}
                    \limits^{\lower 1ex \hbox{$\scriptstyle #1$}} }}
\def\w#1{\widehat{#1}}	
\catcode`\@=11
\def\mymatrix#1{\null\,\vcenter{\normalbaselines\m@th
  \ialign{\hfil$##$&&\quad\hfil$##$\crcr
   \mathstrut\crcr\noalign{\kern-\baselineskip}
   #1\crcr\mathstrut\crcr\noalign{\kern-\baselineskip}}}\,}
\catcode`\@=12
\def\Matrix#1{{\left[\mymatrix{#1}\right]}}

\def\del{\partial}
\def\IR{{\mathbb R}} \def\IC{{\mathbb C}}
 \def\IZ{{\mathbb Z}}
\def\id{1\kern-.35em1} 
\def\half{{ 1 \over 2}}
\def\cA{{\cal{A}}}   
 \def\cF{{\cal{F}}} \def\cG{{\cal{G}}} \def\cH{{\cal{H}}}
 \def\cJ{{\cal{J}}}  \def\cL{{\cal{L}}}
\def\cM{{\cal{M}}} \def\cN{{\cal{N}}}  
  \def\cS{{\cal{S}}}

\def\op{\mathrm}

\def\Lie{{ \op{ Lie} }}  
  \def\Tr{{  \op{Tr  } }}
  
 \def\Ker{{ \op{ Ker} }} 
  \def\Met{{ \op{Met}  }}
     \def\vol{{\op{vol}}}
 \def\sign{{\op{sign} }}
\def\adp{{\mathfrak g}}     
\def\wedgeinner{{\rfloor}}  
\newcommand{\pert}{{\mathop{\mathrm{pert}}\nolimits}} 
\def\PS{\op{PS}}       	

\newtheorem{theorem}{Theorem}[section]

\catcode`\@=11
\def\@optitalic[#1]{\def\@seatemp{#1} \par\noindent{\it \@seatemp}. }
\def\@optit#1{\@ifnextchar [{\@optitalic}{\@optitalic[#1]}}
\newenvironment{remark}{\@optit{Remark}}{\smallskip}
\catcode`\@=12

\def\eqref#1{(\ref{eq:#1})} 
\def\eqdef#1{\ifx#1\end\else\label{eq:#1}\fi}

\def\ft#1{$<$#1$>$} 

\catcode`\@=11 \def\@date{}\catcode`\@=12

\begin{document}
\begin{titlepage}
  \title{Overview and Warmup Example for Perturbation Theory with Instantons}
\author{
                                %
    Scott Axelrod      \\
    Mathematics Department \\
    Massachusetts Institute of Technology \\
    }
\end{titlepage}
\maketitle

\begin{abstract}
The large $k$ asymptotics (perturbation series) for integrals of the
form $\int_{\cal F}\mu e^{i k S}$, where $\mu$ is a smooth top form
  and $S$
is a smooth function on a manifold ${\cal F}$, both of which are
invariant under
the action of a symmetry group ${\cal G}$, may be computed using the stationary
phase approximation.  This perturbation series can be
 expressed as the
integral of a top form on the space $\cM$ of critical points of $S$ mod
the action of ${\cal G}$.
In this paper we overview a formulation of the
``Feynman rules'' computing this top form and a proof that the perturbation
series one obtains is independent of the choice of metric on
${\cal F}$
needed to define it.  We also overview how this definition can
be adapted to the context of $3$-dimensional Chern--Simons quantum
field theory where ${\cal F}$ is infinite dimensional.
This results in a construction of new differential
invariants depending on a closed, oriented $3$-manifold $M$
together with a choice of
smooth component of the moduli space of flat connections on $M$ with
compact structure group $G$.
To make this paper more accessible we warm up with a trivial example and
only give an outline of the proof that one obtains invariants in the
Chern--Simons case.  Full details will appear elsewhere.
\end{abstract}

\catcode`\@=11
\renewcommand{\@makefnmark}{}
\footnotetext{
      This work was supported in part by the Divisions of Applied
      Mathematics of the U.~S.~Department of Energy under contract
      DE-FG02-88ER25066 as well as by an Alfred P. Sloan
      Research Fellowship.}
\renewcommand{\@makefnmark}{\mbox{$^{\@thefnmark}$}}
\catcode`\@=12

\section{Introduction}

In previous joint work with I.M.~Singer \cite{ASI,ASII}, we defined
Chern--Simons perturbation theory about an acyclic flat connection on some
principle bundle $P$ with compact structure group $G$ over a closed, oriented
$3$-manifold $M$. In my talks at the conference I explained how to generalize
this result to  perturbation theory about a smooth component $\cM$ of the
moduli space of flat connections.  Specifically, I proved the following
theorem.

\begin{theorem}
  Let $\cM$ be a smooth component of the moduli space of
  flat connections on $P \rightarrow M$.
  Then Chern-Simons perturbation theory about $\cM$ defines a
  differential invariant
\begin{equation}
  Z^{\mathrm{pert}}_k (M, {\cal M}) \in PS(k)
\end{equation}
  depending on the choice of $M$ and $\cM$.
\end{theorem}

\begin{remark}[Remark 1]
  Let $\cS$ be the set of all flat connections on $P$ representing
  elements of $\cM$.
  Our smoothness assumption on $\cM$ includes the requirement that
  $\cS\to\cM$ be a smooth bundle (with fibers a homogeneous space $\cG/\cH$)
and that the tangent space $T_{A_0}\cS$ to
$\cS$ at a point $A_0$ be equal to the deformation tangent space,
  that is, the kernel of the exterior derivative operator acting
on the space $\Omega^1(M;\adp)$ of $1$-forms on $M$ with values in
the adjoint bundle associated to $P$.
\end{remark}
\begin{remark}[Remark 2]
  The ring, $PS(k)$, of perturbation series is the ring generated
  by formal power series in $k^{-1}$, fractional powers of $k$,
  and oscillatory exponentials in $k$ of the form $e^{i k S_0}$.
  Since $k$ only takes on integer values, $S_0$ is an element of
  $\IR/2\pi\IZ$.
\end{remark}

Let us explain our strategy for arriving at a definition
of ``Chern--Simons perturbation theory about $\cM$''.  To begin,
we consider a finite dimensional integral of the form
\begin{equation}
  Z_k = {1\over\vol(\cG)}\int_{A\in\cF} \mu e^{ikS(A)}
\eqdef{pathint}\end{equation}
where $S:\cF\to\IR$ is a Morse-Bott function, and $\mu$ is a smooth measure
on a manifold $\cF$, both of which are invariant under the action
of a compact symmetry group $\cG$ with a bi-invariant volume form of
total volume $\vol(\cG)$.   Using a partition of unity, we can break up
the above integral into a sum of contributions from the integral near
each of the components of the critical point set of $S$ and a
contribution from the integral away from the critical point set.  The
large $k$ asympotics of the latter vanish faster than any power of
$k$.  The large $k$ asymptotics of the integral \eqref{pathint}\ near a
component $\cS$ of the critical point set is what we mean by
``perturbation theory'' about $\cS$.  Since we can also think of the
integral as an integral over $\cF/\cG$, we can equally well speak of this as
perturbation theory about the quotient space $\cM=\cS/\cG$.  For this
finite dimensional integral, the stationary phase approximation yields
an algorithm to calculate perturbation theory near $\cS$.
For simplicity, we assume $\cF$, $\cS$, and $\cM$ are oriented so that
a smooth measure on any of these spaces is just a smooth top form.

The challenge is to formulate the stationary phase approximation
computing the large $k$ asymptotics of \eqref{pathint}\ in a way such that
the result can be generalized to the case of Chern--Simons theory where
the space $\cF$ becomes the infinite dimensional space of connections
on a principal bundle $P$, the group $\cG$ becomes the group of base
preserving automorphisms of $P$, the Morse-Bott function $S$ becomes
the Chern--Simons functional, $\cM$ is a smooth component of the moduli
space of flat connections on $P$, and $k$ is a positive integer called
the level of the theory.  The generalization of the stationary phase
algorithm to the field theory case involves integrals over multiple
copies of the spacetime manifold $M$.  In most quantum field theories
these integrals diverge and one has to perform the procedure of
perturbative regularization and renormalization.  Fortunately, just as
in \cite{ASI}, it is possible to package the finite dimensional answer
in such a way that its generalization to the Chern-Simons case involves
only convergent integrals.  (For experts, there is a single
point-splitting regularization one must introduce for the
``tad-pole''.) To formulate a concrete algorithm to compute the
perturbation theory in the case of a finite dimensional integral, we
need to pick a metric $g$ in the space $\Met$ of $\cG$ invariant
metrics on $\cF$.  The result, however, is {\it a priori} independent
of $g$ because the integral \eqref{pathint}\ that one is
calculating the asymptotics of does not depend on $g$.  In the
Chern--Simons case, the space $\Met$ will denote the space of
Riemannian metrics on $M$, each of which determines a $\cG$ invariant
metric on the space $\cF$ of connections.  The difficulty is that there
is no {\it a priori} proof of the independence of $g$ in the Chern--Simons
case since the functional integral is ill-defined.  What we do is
formulate not only the algorithm to compute the stationary phase
approximation but also a proof of independence of $g$ in the finite
dimensional case in a way that can be carried over to the quantum
field theory problem.

To calculate the stationary phase approximation in the finite
dimensional case what one has to do is to first integrate over the
directions normal to the set $\cS$ of critical points of $S$ and then
to subsequently integrate over $\cS$.   The standard method of
stationary phase about an isolated critical point (as for example
captured in the Feynman rules described in the Appendix)
 applies to the integral in the normal directions.  To do this properly
 however, one needs to carefully keep track of what happens to the
 differential forms as one makes the split between the normal
 directions and the directions tangent to $\cS$. We pick a metric
 $g\in\Met$ in order to identify ``normal directions'' with directions
 orthogonal to the tangent space of $\cS$.  To have a mechanism to keep
track of the dependence on the metric, we
 introduce the evaluation map
\begin{equation}
  E : N\times\Met \to \cF
.\end{equation}
Here $N\to\cS$ is the normal bundle to $\cS$ in $\cF$.  The fiber of
the bundle $N\times\Met\to\cS\times\Met$ above a point $(A_0,g)$ is
just the orthocomplement of the subspace $T_{A_0}\cS$ in $T_{A_0}\cF$
with respect to the metric $g$.  The evaluation map takes an element
$B$ of this space and sends it to the point determined by the
exponential map coming from the geodesic flow from $g$, i.e.
\begin{equation}
  E((A_0,g),B) = \exp_{A_0}(B)
.\end{equation}
For brevity, we will write $\w N$ for $N\times\Met$, $\w\cS$ for
$\cS\times\Met$, and $\w\cM$ for $\cM\times\Met$.  In general, objects adorned
with a hat fiber over $\Met$ or depend on a choice of $g\in\Met$.

In addition, to get a formula that will be well defined and not involve
infinite dimensional integrals in the Chern--Simons example, we must
make use of invariance under the group $\cG$ to replace the ill-defined
integral over the infinite dimensional space $\cS$ by a well-defined
integral over the finite dimensional space $\cM=\cS/\cG$ of gauge
equivalence classes of flat connections.

So far we have stated the basic conceptual idea.  There are quite a few
technical details where we generalize some standard tricks of
supermanifold theory and gauge fixing theory in order to come up with
an explicit formula for the perturbation series.  For the main result,
the definition can be stated in terms of the structures on the
deformation complex for the calculation of $\cM$ that arise because we
have metrics present and so may define Hodge theory and we have the
Taylor series for the action $S$ near $\cM$.  Even though we expended a
lot of effort in order to find the ``right formulation'' of the
definition, it is rather technical.
To make this paper accessible to non-experts, we will
focus our attention in \S 2 to a trivial example which
illustrates some of the important points that one needs to understand
in formulating the general perturbaton theory.
A more detailed overview of the general finite dimensional construction
and the application to Chern--Simons theory is given in \S 3 and
\S 4, but the complete details  will appear elsewhere \cite{AXE}.
An appendix reviewing the derivation of the Feynman rules and a list of
notations are included at the end for the reader's convenience.

\section{Trivial Example}

For our trivial example we take $\cF$ to be $\IR^2$
with the usual $(x,y)$ coordinates, $\cG$ to be the trivial group, $\mu$ to
be the standard area form $dx\wedge dy$, and the Morse-Bott function $S$ to
have the form
\begin{equation}
  S(x,y) = S_0 +\half H(x) y^2 + {1\over 3!} V_3(x) y^3 +
          {1\over 4!} V_4(x) y^4
.\end{equation}
Hear $S_0$ is a constant, $H(x)$ is any positive function of $x$ and we
assume for simplicity that the Taylor series with respect to $y$
stops at the fourth order term.  The positivity of $H(x)$ insures that
the real axis $\cS = \{(x_0,0); x_0\in\IR\}$ is a smooth component of
the critical point set.  For our purposes, we will just assume that
there are no other components.  Since we are assuming $\cG$ is trivial,
the ``moduli space'' $\cM$ is the same as $\cS$.
To be honest, we should note that the integral \eqref{pathint}\ does not
actually converge.  To make it converge we should replace $\cF$ by
$S^1\times S^1$ or multiply $dx\wedge dy$ by a bump function supported near
the origin, but we won't introduce notation for this.

The stationary phase approximation for this example is of course just the
integral over $\cS$ of the result obtained by doing the stationary phase
approximation to the integral over $y$.  Abstractly, one can think of the
integral over $y$ as the integral over the normal directions to $\cS$ with
respect to the standard metric on $\IR^2$.  Our goal is to see very
explicitly why one gets the same result if one uses a different metric.
To keep things simple, we take $\Met=\IR$ and identify $g\in\Met$ with the
translationally invariant metric
represented by the matrix $\Matrix{1& -g \cr -g & 1}$.  The normal
directions to $\cS$ are then lines with slope $g$ relative to the $y$
axis.  Note that the normal bundle to $\cS$ is trivial (as one would
expect in such a trivial example). We may identify the total space of
$\w N$ with $\IR^3$, with coordinates $(x_0,g,t)$; $x_0$ determines a
point $A_0=(x_0,0)$ in $\cS$; $g$ is a point in $\Met$; and $t$
is the coordinate on the fibers of $\w N$.  Since the exponential map is
linear, we have
\begin{equation}
    (x,y) = E(x_0,g,t) = (x_0,0) + t( g, 1),\qquad{t\in\IR}
\eqdef{lcvar}\end{equation}
going through $(x_0,0)$ with slope $g$.
Our setup is depicted in Figure 1.

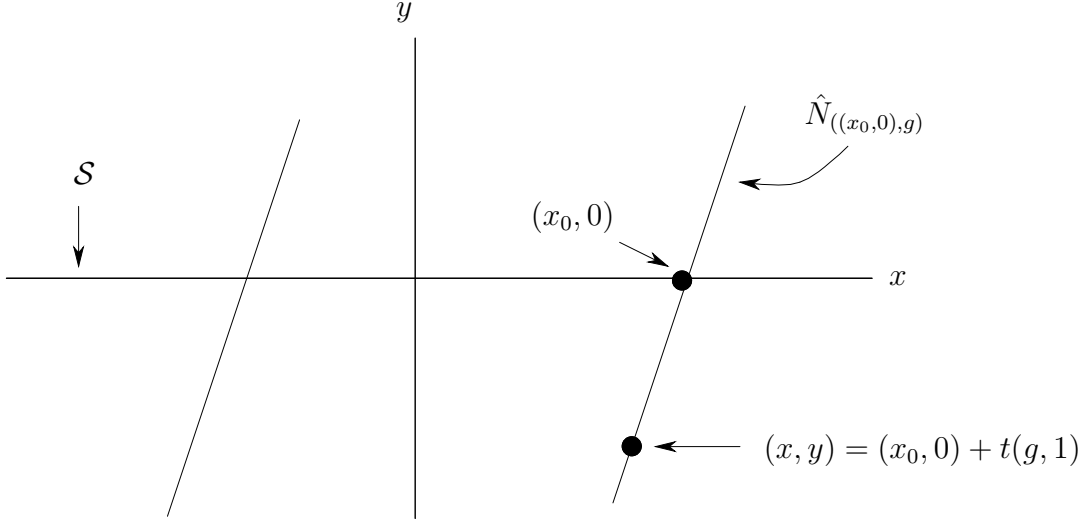
\begin{figure}
\setlength{\unitlength}{0.00083333in}
\begingroup\makeatletter\ifx\SetFigFont\undefined%
\gdef\SetFigFont#1#2#3#4#5{%
  \reset@font\fontsize{#1}{#2pt}%
  \fontfamily{#3}\fontseries{#4}\fontshape{#5}%
  \selectfont}%
\fi\endgroup%
{\renewcommand{\dashlinestretch}{30}
\begin{picture}(6751,3294)(0,-10)
\put(4248,1497){\blacken\ellipse{120}{120}}
\put(4248,1497){\ellipse{120}{120}}
\put(3933,462){\blacken\ellipse{120}{120}}
\put(3933,462){\ellipse{120}{120}}
\path(483,1962)(483,1587)
\blacken\path(453.000,1707.000)(483.000,1587.000)(513.000,1707.000)(483.000,1671.000)(453.000,1707.000)
\path(3858,1737)(4158,1587)
\blacken\path(4037.252,1613.833)(4158.000,1587.000)(4064.085,1667.498)(4082.868,1624.566)(4037.252,1613.833)
\path(4608,462)(4083,462)
\blacken\path(4203.000,492.000)(4083.000,462.000)(4203.000,432.000)(4167.000,462.000)(4203.000,492.000)
\texture{55888888 88555555 5522a222 a2555555 55888888 88555555 552a2a2a 2a555555 
	55888888 88555555 55a222a2 22555555 55888888 88555555 552a2a2a 2a555555 
	55888888 88555555 5522a222 a2555555 55888888 88555555 552a2a2a 2a555555 
	55888888 88555555 55a222a2 22555555 55888888 88555555 552a2a2a 2a555555 }
\path(5433,1512)(33,1512)
\path(5433,1512)(33,1512)
\path(3816,111)(4641,2586)
\path(1038,27)(1863,2502)
\path(2583,3012)(2583,12)
\path(2583,3012)(2583,12)
\path(5283,2337)	(5227.924,2286.354)
	(5179.628,2243.460)
	(5137.232,2207.661)
	(5099.858,2178.296)
	(5036.659,2136.234)
	(4983.000,2112.000)

\path(4983,2112)	(4925.533,2099.192)
	(4890.475,2095.990)
	(4849.736,2094.922)
	(4802.218,2095.990)
	(4746.822,2099.192)
	(4682.449,2104.529)
	(4608.000,2112.000)

\blacken\path(4730.510,2129.071)(4608.000,2112.000)(4724.130,2069.411)(4691.524,2103.069)(4730.510,2129.071)
\put(5538,1467){\makebox(0,0)[lb]{\smash{{{\SetFigFont{12}{14.4}{\rmdefault}{\mddefault}{\updefault}$x$}}}}}
\put(5013,2472){\makebox(0,0)[lb]{\smash{{{\SetFigFont{12}{14.4}{\rmdefault}{\mddefault}{\updefault}$\hat N_{((x_0,0), g)}$}}}}}
\put(4758,387){\makebox(0,0)[lb]{\smash{{{\SetFigFont{12}{14.4}{\rmdefault}{\mddefault}{\updefault}$(x, y) = (x_0, 0) + t (g, 1)$}}}}}
\put(2463,3147){\makebox(0,0)[lb]{\smash{{{\SetFigFont{12}{14.4}{\rmdefault}{\mddefault}{\updefault}$y$}}}}}
\put(3303,1842){\makebox(0,0)[lb]{\smash{{{\SetFigFont{12}{14.4}{\rmdefault}{\mddefault}{\updefault}$(x_0, 0)$}}}}}
\put(453,2097){\makebox(0,0)[lb]{\smash{{{\SetFigFont{12}{14.4}{\rmdefault}{\mddefault}{\updefault}$\cal S$}}}}}
\end{picture}
}
\caption{Setup for Trivial Example}
\end{figure}

We may now calculate.
\begin{equation}
\renewcommand{\arraystretch}{2.5}
\begin{array}{rcl}
    E^*(S) (x_0,g,t)  =  S(x_0+tg,t)
      & = & S_0 + H (x_0){ t^2 \over 2}
\\
      &   & + \left( 3 g  H' (x_0) + V_3 (x_0) \right) {t^3\over 3!}
    + \cdots \\
    E^*(\mu)(x_0,g,t) = E^*(dx \wedge dy)
    & = & \left[ d x_0 + t\, d g + g\, d t \right]\, \wedge d t\\
    & = & \left[ 1 + t\, d g\, I
                 \left( \frac{\partial}{\partial x_0} \right)
          \right] \wedgeinner (d x_0 \wedge d t)
.\end{array}
\eqdef{estar}\end{equation}
Note that the quantity in brackets in the last line has both an
operator of exterior product with $dg$ and an interior product operator,
$I(\part{x_0})$,
in the direction of $\part{x_0}$.  If there where several interior
product operators (as there would be in a higher dimensional example),
we would write all of them on the right and view them as a multivector,
i.e.  as an element of $\Lambda^*(T\cS)$.  Accordingly we view the
quantity in brackets as a form with values in $\Lambda^*(T\cS)$.  The
symbol $\wedgeinner$ in the last line symbolizes that a form with
values in multi-vectors is acting by a combination of wedge product and
inner product.

To calculate the stationary phase approximation to $Z_k^\pert$ we must first
hold $g\in\Met$ and $A_0=(x_0,0)\in\cS$ fixed and calculate the stationary
phase approximation to the integral over the normal directions (i.e. over $t$).
 This gives us a quantity which we write
\begin{equation}
\hat Z^\pert_k = \int^\pert_{t \in \IR} E^*(\mu \exp i k S)(x_0,g,t)
    \in \Omega^1_{cl}(\cS\times\Met)\otimes\PS(k)
.\end{equation}
The symbol $\int^\pert$ means that we are not to actually evaluate the
integral, but merely to calculate its stationary phase approximation.
This is given explicity by the Feynman rules which are briefly
recalled in the appendix.
Notice that  $\hat Z_k^\pert$ is a closed form on $\cS\times\Met$
(which is denoted by the subscript $cl$ on $\Omega^1_{cl}$).
This
follows because it is a (perturbative) push-forward integral of a closed form.
$Z_k^\pert$ is the integral of this quantity over $\cS$,
\begin{equation}
Z_k^\pert = \int_{A_0=(x_0,0)\in\cS} \hat Z_k^\pert
         \in \Omega^0_{cl}(\Met)\otimes\PS(k) = \PS(k)
.\end{equation}
Since $\w Z_k^\pert$ was a closed $1$-form on $\cS\times\Met$,
$Z_k^\pert$ is a closed $0$-form on $\Met$, i.e. it is metric independent.
Of course, in this finite dimensional example, $Z_k^\pert$ actually
is the asymptotic series for $Z_k$ and so is {\it a priori} independent of
the metric.  In the case of Chern--Simons field theory, however,
to prove metric independence, we need to give a direct proof that
$\w Z_k^\pert$ is closed.  As a warmup, we will verify explicitly
$\w Z_k^\pert$ is closed to subleading order for the trivial
example.

Applying the result in Appendix A, carrying all of the
operators on forms along for the ride, we find
\begin{eqnarray}
 \hat Z^\pert_k  & = & \hat Z^{hl}_k\, \wedgeinner\, \hat Z^{sc}_k
\eqdef{tz}\\
   \hat Z^{sc}_k & = &
       \sqrt{\frac{2 \pi i}{k}}
          \frac{e^{i k S_0}}{\sqrt{H(x_0)}}\, d x_0
       \in\Omega^1_{cl}(\cS)\otimes\PS(k)
         \subset\Omega^1_{cl}(\w\cS)\otimes\PS(k)
\eqdef{tzsc}\\
    \hat Z^{hl}_k & = &
    \sum_{\mathrm{graph}\, \Gamma} \frac{1}{S (\Gamma)} I
    (\Gamma)
     \in\left[\mathop{\oplus}\limits_l\Omega^l(\w\cS;\Lambda^l(T\cS))
        \right]  \otimes\IC[[{1\over k}]]
\eqdef{tzhl}
\end{eqnarray}
The leading order ``semi-classical'' term $\hat Z_k^{sc}$ is not only
closed, but manifestly metric independent.  This property carries
through (when appropriate counterterms are added) to the case of
quantum field theory.
The Feynman rules for the graphs contributing to
$\hat Z_k^{hl}$ are as follows:

\begin{eqnarray}
  \epsfbox{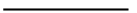}
  & & -\frac{1}{i k} H (x_0)^{-1} \\
  \lower .25in\hbox{\epsfbox{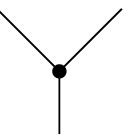}}
  && i k
  \left.
    \frac{\partial^3 S}{\partial t^3}
  \right|_{x_0} = i k
  \left[
    3 H' (x_0) g + V_3 (x_0)
  \right] \\
  \lower .25in \hbox{\epsfbox{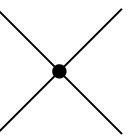}}
  && i k
  \left.
    \frac{\partial^4 S}{\partial t^4}
  \right|_{x_0} = ik
  \left[ 6 H''(x_0) g^2 + 4 V'_3(x_0) g + V_4(x_0) \right]\\
  \vdots \hspace*{.25in}  && \\
  \epsfbox{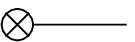}
  &&dg\,  I
  \left(
    \frac{\partial}{\partial x^0}
  \right) \eqdef{extgraph}
\end{eqnarray}

We consider the vertex in  \eqref{extgraph}\ an
external vertex and only allow it to appear at most once in any
given graph because it comes from the term $t\, dg\, I(\part{x_0})$
in $E^*(\mu)$ (see \eqref{estar}) rather than from a term in the Taylor
expansion in $S$.

We may expand the ``higher loop'' term $\w Z_k^{hl}$ in powers of $k$:
\begin{equation}
     \w Z_k^{hl} = 1 + \sum_{l > 1} \w I_l (-i k)^{1 - l}
\end{equation}
where $(-ik)^{1-l} \w I_l$ is a sum of $I(\Gamma)/S(\Gamma)$ over all graphs
$\Gamma$ with
\begin{equation}
 \mbox{number of edges} - \mbox{number of internal vertices} = l-1
.\end{equation}
The graphs $\Gamma$ contributing to $\w I_l$ are the graphs
such that
\begin{equation}
b_1(\Gamma) + \mbox{number of external vertices} =
   l + (b_0(\Gamma) -1)
,\end{equation}
where $b_0(\Gamma)$ and $b_1(\Gamma)$ are the Betti numbers of $\Gamma$
(the number of components and the number of independent loops, resp.).
Equivalently, a graph $\Gamma$ contributes to
$\w I_l$ if the graph obtained from $\Gamma$ by adding a loop at each external
vertex has $l+(b_0(\Gamma)-1)$ loops.   We refer to $\w I_l$ as the $l$'th
order or $l$-loop piece of $\w Z_k^{hl}$.

The subleading term $\w I_2$ is given by
\begin{equation}
    (-i k)^{-1} \w I_2 = \frac{1}{12}\:
    \lower .10in \hbox{\epsfbox{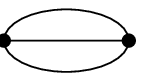}} +
    \frac 18\: \lower .10in \hbox{\epsfbox{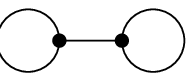}} +
    \frac 18\: \lower .10in \hbox{\epsfbox{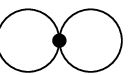}} +
    \frac 12\: \lower .10in \hbox{\epsfbox{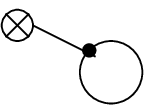}}
\eqdef{itwo}\end{equation}
Here the picture of a graph $\Gamma$ represents the term $I(\Gamma)$.
Evaluating this, we obtain
\begin{eqnarray}
 \w I_2   &=&
    \left[\frac{1}{12} + \frac 18\right] H^{-3} (3 H' g + V_3)^2
      -\frac 18 H^{-2}(6 H'' g^2 + 4 V_3' g + V_4)
\cr
  & & -\frac 12 H^{-2} (3 H' g + V_3)\, dg\, I(\part{x_0})
\end{eqnarray}

Finally, the subleading contribution to $\w Z_k^\pert$ equals
the $k$-dependent term  ${e^{ik S_0}\over ik}\sqrt{2\pi i\over k}$ times
the following:
\begin{eqnarray}
 \hat I_2 \wedgeinner \frac{d x_0}{\sqrt{H}}
   &=&
    \left(
         \left[\frac{1}{12} + \frac 18\right] H^{-\frac72} V_3^2
        -            \frac 18                 H^{- \frac 52} V_4
    \right) dx_0
\cr
   & & +\frac 18\left[d x_0 \part{x_0} + d g \part{g}\right]
    \left(
      - 6 H^{- \frac 52} H' g^2 - 4 H^{- \frac 52} V_3 g
    \right)
\eqdef{trans}
\end{eqnarray}

The first term on the right in \eqref{trans}\ is a closed form
and the second term is exact on $\cS\times\Met$.
Thus the final result is a closed form on $\cS\times\Met$ as desired.
We could, of course, have obtained the first term (which is the only term
which makes a contribution to $Z_k^\pert$) much more easily by only
considering the usual metric on $\cF=\IR^2$ with slope $g$ equal to $0$.
The point to notice, however, is that the direct verification that
$\w Z_k^\pert$ is closed, even at the lowest subleading order in this
most trivial example, involves a delicate cancellation among
several terms.
In particular notice that it is only after summing over graphs with varying
numbers of vertices that we obtain a closed form.

\section{Outline for the General Finite Dimensional Case}

In this section we give a very sketchy outline of our formulation of
the perturbation series about a component $\cS$ of the critical point set of
$S$ for a general integral of the form
\eqref{pathint}.  We assume that $\cF$ is an affine space, that $\cG$ acts
affinely, $S$ is a cubic function, and that $\Met$ consists of translationally
invariant metrics, as is the case for Chern--Simons theory.

\subsection{Strategy in Language Using Differential Forms}

The desired perturbation series may be expressed as an integral over $\cM$,
\begin{equation}
 Z_k^\pert(\cM) = \int_{\cM} \hat Z_k^\pert
   \in\Omega^0_{cl}(\Met)\otimes\PS(k) = \PS(k)
.\end{equation}
$Z_k^\pert(\cM)$ is a closed $0$-form on $\Met$ (and hence a constant)
because the integrand here is a closed form on $\cM\times\Met$,
\begin{equation}
\hat Z_k^\pert\in\Omega_{cl}^{|\cM|}(\cM\times\Met)\otimes\PS(k)
.\end{equation}
Note that we denote the dimension of the manifold $X$ by $|X|$.
Our goal is to find an explicit formula for $\hat Z_k^\pert$ and
a proof that it is closed which will carry over to the case of
Chern--Simons quantum field theory.

$\hat Z_k^\pert$ is the image of the integrand $\mu e^{i k S}$ of
\eqref{pathint}\ under the following sequence of maps:
\begin{eqnarray}
\eqdef{ralph}
 \Omega^{|\cF|}_{cl}(\cF)^\cG
  &\maplongright{E^*}&
 \Omega^{|\cF|}_{cl}(\w N)^\cG
   \ \maplongright{I_{\nabla^{\hat N \rightarrow \hat{\cS}}}}\
 \Omega^{|\cS|}_{cl}(\w\cS; \Omega^{|\cF|-|\cS|}_{vert}(\w N) )^\cG
\nonumber\\
\nonumber\\
  &\maplongright{\int^\pert_{\hat N \rightarrow \hat\cS}}&
 \Omega^{|{\cS}|}_{cl}(\w\cS)^\cG\otimes\PS(k)
   \maplongright{\left[\vol(\cH)\mu_{\cG/\cH}\right]^{-1}}
 \Omega^{|\cM|}_{cl}(\w\cS)_{\mathrm{basic}} \otimes\PS(k)
\nonumber\\
\nonumber\\
  &\maplongright{\cong} &
 \Omega^{|\cM|}_{cl}(\w\cM)\otimes\PS(k)
\end{eqnarray}
The first map, $E^*$, is just pull back under the evaluation map.
To define the second map we must exhibit a connection on
the vector bundle $\w N\to\w\cS$ (the normal bundle crossed with $\Met$),
which we view as a
complement $T_{hor}\w N$ to $T_{vert}\w N$ within $T\w N$.
To handle the group theory in a way that will gives us a sensible
answer in the
field theory problem (which might be interpretted by physicists as
a generalization of a familiar method of  ``gauge fixing''), we require that
directions  along the group orbits on $\w N$ be horizontal\footnote{
  More precisely, if there is an isotropy group
  $\cH$, we require the direction generated by elements of $\Lie(\cG)$
  orthogonal to $\Lie(\cH)$ be horizontal.}.
The remaining horizontal direction for the connection
$\nabla^{\w N\to\w\cS}$ are defined using the fact that
$\w N\to\w\cS$ is a subbundle of the Riemannian bundle with
connection $T\cF|_\cS\times\Met\to\w\cS$.
The map $I_{\nabla^{\w N\to\w\cS}}$ comes from the identification
\begin{equation}
  \Lambda^*(T^*\w N)=
  \Lambda^*(T^*_{hor}\w N)\otimes \Lambda^*(T^*_{vert}\w N)
.\end{equation}
The third map, $\int^\pert_{\w N\to\w S}$, in \eqref{ralph}\ is just the
``perturbative integral'', or stationary phase approximation,
to the integral over the fibers of $\w N\to\w\cS$.
This may be calculated using the Feynman rules as desribed in
Appendix A since the Morse-Bott
condition on $S$ insures that the zero vector is an isolated critical point of
the restriction of $S$ to a given fiber of $\w N\to\w\cS$.
The fourth map, $\left[\vol(\cH)\mu_{\cG/\cH}\right]^{-1}$, is just division by
the volume form along the group orbits (which is normalized so that it's
integral over an orbit is $\vol(\cG)$).
A precise formulation of what is meant by this division requires
 the use of the natural connection
$\nabla^{\w\cS\to\w\cM}$
on $\w\cS\to\w\cM$.
(Then the volume form $\mu_{\cG/\cH}$ on the coset space
  $\cG/\cH$, which is diffeomorphic to the group orbits,
  determines a vertical differential form on $\w\cS$ which is what
  we actually divide by.)
The superscript $\cG$ throughout \eqref{ralph}\ indicates that we
are working with $\cG$-invariant subspaces.  The subscript
``$\mathrm{basic}$'' on $\Omega^*(\w\cS)$ in the space appearing
before the final arrow indicates that the forms
are both $\cG$-invariant and annihilated by the interior product
operator with any vector field along the group orbits.  The
isomorphism in the final map in \eqref{ralph}\ is the inverse of the
pullback map associated to the projection from $\w\cS$ to $\w\cM$.

\subsection{\bf Explicit Formulas Obtained Using Supervariables}

To obtain an explicit formula embodying the sequence \eqref{ralph}\ of maps, we
introduce some supervariables.  We let $A$ denote a variable in
$\cF$, and introduce Fermionic variables $\chi$
valued in $T^*\cF_A$ and $\delta A$ valued in $T\cF_A$.
We may identify a function of $\delta A$ with an element of
$\Lambda^*(T^*\cF_A)$ and a function of $\chi$ with an element of
$\Lambda^*(T\cF_A)$.  We will not review the theory of
supermanifolds\cite{DEWITT} for readers who are not familiar,
but simply present the following Rosetta stone which will suffice for
our purposes:
\begin{eqnarray}
   \int_{A\in\cF} \mu e^{i k S}
   & = &
   \int_{A \in \cF} \int_{\chi \in \left[T^*\cF_A\right]_-}
        \int_{\delta A \in \left[T\cF_A\right]_-}
         e^{i<\chi, \delta A> + i k S (A)}
   \eqdef{pif}\\
  \Omega^* ({\cal F};\wedge^* (T {\cal F} ))
  & = &
  C^\infty ( \{ A_0, \chi, \delta A \} )
  \\
  d x^\mu &\longleftrightarrow& \delta A^\mu
  \\
  I(\part{A^\mu}) &\longleftrightarrow& \chi_\mu
  \\
  \Omega^1(\cF;\wedge^1(T\cF))\ni \id
     &\longleftrightarrow&  <\chi,\delta A>
\end{eqnarray}
The minus subscripts on $\left[T^*\cF_A\right]_-$ and
$\left[T\cF_A\right]_-$ are just there to remind us that the variables
taking values in those space are Fermionic.
This integral is manifestly independent of the choice of $g\in\Met$,
but for later purposes we should consider this as one of our variables
and also introduce a Fermionic variable $\delta g\in T\Met_g$ so that
function of the pair $(g,\delta g)$ correspond to forms on $\Met$.

The image of $\mu e^{ikS}$ under \eqref{ralph}\ is calculated using
a sequence of changes of supervariables and (perturbative) fiber integrals.  To
define these, we first define the deformation chain
complex associated to the ``moduli problem'' of
calculating $\cM$ and its tangent space.
At a point $A_0\in\cS$, it is given by:
\begin{equation}
  \begin{array}{c@{}c@{}c@{}c@{}c@{}c@{}c@{}c@{}c@{}c@{}c}
    0 \,& \longrightarrow            \,& \mathrm{Lie}(\cG)
      \,& \maplongright{-T_{A_0}}    \,& T\cF_{A_0}
      \,& \maplongright{H(A_0)}      \,& T^*\cF_{A_0}
      \,& \maplongright{-(T_{A_0})^T}\,& \mathrm{Lie}(\cG)^*
      & \longrightarrow            &  0 \\*[4pt]
      && \| && \| && \| && \| \\*[4pt]
    0 &\longrightarrow & \Omega^0_M &
    \maplongright{D_{A_0}^0} & \Omega^1_M &
    \maplongright{D_{A_0}^1} & \Omega^2_M &
    \maplongright{D_{A_0}^2} & \Omega^3_M & \longrightarrow & 0
  .\end{array}
\eqdef{defcplx}\end{equation}
Here $T_{A_0}$ is the infinitesimal action of the group $\cG$ at the point
$A_0$ and  $H(A_0)$ is the Hessian of $S$ at $A_0$.
Diagram \eqref{defcplx}\ is a definition of the complex $\Omega^*_M$ and
the differential $D^*_{A_0}$.
Note that the notation for the generic finite dimensional situation we
are currently considering is also ideally suited for the Chern--Simons
quantum field theory we will be considering.
These complexes fit
together to form a bundle $\w\Omega^*_M$ of chain complexes over
 $\w\cS$ (which is trivial in the $\Met$ directions).  In fact, using
the metric, the connections $\nabla^{\w\cN\to\w\cS}$ and
$\nabla^{\w\cS\to\w\cM}$, and the Taylor expansion of the function $S$ near
$\cS$, $\w\Omega^*_M$ becomes a bundle of exterior algebras with Hodge
decomposition and connection, which are all invariant under an action of $\cG$
which lifts the action of $\cG$ on $\cS$.  (There is also a product structure
which satisfies a variant of the rules for the product in a  differential
graded algebra)
The Hodge structure means that we can decompose $\w\Omega^*_M$ as
a direct sum of a piece $\w\Omega^*_d$ which equals the image of $D^*$, a
piece $\w\Omega^*_\delta$ which is the image of the adjoint of $D^*$, and a
piece $\w\Omega^*_h$ which is the orthocomplement of the other two:
\begin{equation}
\w\Omega^*_M = \w\Omega^*_h\oplus\w\Omega^*_d\oplus\w\Omega^*_\delta
.\end{equation}
The name $\Omega^*_M$ was chosen, because in the case of
Chern--Simons theory
\begin{equation}
\Omega^*_M = \Omega^*(M;\adp)
\end{equation}
and $D^*_{A_0}$ is just the exterior derivative twisted by $A_0$.
The Hodge structure in the Chern--Simons case is the familiar one from Hodge
theory of differential forms, and the product structure is a combination of
wedge product and Lie bracket.
In general, we write the product as a bracket operation
$[\cdot,\cdot]:\Omega^j_M\otimes\Omega^k_M\to\Omega^{j+k}_M$.
The bracket with $0$ forms comes from the Lie algebra action.
The bracket on one forms $B\in T\cF_{A_0}=[\w\Omega^1_M]_{(A_0,g)}$
is determined by the cubic term in $S$ so that
\begin{equation}
S(A_0+B) = S(A_0) + \half <B, H(A_0) B> + {1\over 6} < B,[B,B] >
.\end{equation}

We remark that all of the above goes through even when $\cF$ is not affine, or
$S$ is not cubic, except then the fibers of  $\w\Omega^*_M\to\w\cS$ would come
equipped with even more structure that ``keeps track'' of the non-linearities.
If we wanted to allow for this, we would have to
add some extra terms below.

We may now define our change of variables.
The variables $g$ and $\delta g$ stay as before.
$A$ is replaced by a point $A_0\in\cS$ and a normal direction
$B\in N_{(A_0,g)}$:
\begin{equation}
  A = A_0 + B
.\eqdef{chone}\end{equation}
Note that
$N_{(A_0,g)}$ just equals $[\w\Omega^1_\delta]_{(A_0,g)}$ since the
normal directions are orthogonal to $T_{A_0}\cS$ and the latter space
is  $\Ker(H(A_0))=\Ker(D^1_{A_0})$.
$\chi$ is
replaced by
its harmonic piece $\chi_h\in[\w\Omega^2_h]_{(A_0,g)}$,
its exact piece    $\chi_d\in[\w\Omega^2_d]_{(A_0,g)}$, and
its coexact piece  $\chi_\delta\in[\w\Omega^2_\delta]_{(A_0,g)}$:
\begin{equation}
 \chi = \chi_h + \chi_d + \chi_\delta
.\eqdef{chtwo}\end{equation}
Finally, $\delta A$ is replaced by the variables
$\delta A_{0,h}$ belonging to $[\w\Omega^1_h]_{(A_0,g)}$,
$c$ belonging to $[\w\Omega^0_h]_{(A_0,g)}$ (the subspace of $\Lie(\cG)$
orthogonal to the isotropy group at $A_0$), and
$\delta_{vert} B$ belonging to $[\w\Omega^1_\delta]_{(A_0,g)}$.
In order for these to combine into the element $\delta A$ behaving as
an element of $T_A\cF$, we set
\begin{equation}
  \delta A = \delta A_{0, h} + T_A(c) + \delta_{vert} B
             +\left(\delta_{(\delta A_{0, h},\delta g)}P_N \right) B
.\eqdef{chthree}\end{equation}
Here $P_N$ is the orthogonal projection operator from
$\w\Omega^1=T\cF$ to $\w N=\w\Omega^1_\delta$, and the
expression $\delta_{(\delta A_{0, h},\delta g)}$ stands for the
covariant derivative operator acting in the direction of
$(\delta A_{0, h},\delta g)$.
The second term in \eqref{chthree}\ is in the direction of the orbit through
$A$,
\begin{equation}
T_A(c) = T_{A_0}(c) + [c,B] = - (D^1(c) + [B,c])
.\end{equation}
In summary, equations \eqref{chone},\eqref{chtwo}, and
\eqref{chthree}\ define a change of variables from the variables on
 the left below
(with $A$ bosonic and $\chi$ and $\delta A$ Fermionic) to the
variables on the right below (with the first group Bosonic and the
last two groups Fermionic):
\begin{equation}
 A,\ \chi,\ \delta A
\quad\longleftrightarrow\quad
 (A_0, B),\  (\chi_h, \chi_d,\chi_\delta),\
 (\delta A_{0,h}, \delta_{vert} B, c )
.\end{equation}

We write \eqref{pif}\ in these new variables, trivially integrate
out the variables $\chi_d$ and $\delta_{vert} B$, and do a perturbative
integral over the combined variable\footnote{
  The variable $\cA$ generalizes the variable of the same name
  appearing in \cite{ASI}.
  }
\begin{equation}
\cA = c + B + \chi_\delta \in [\w\Omega^*_\delta]_{(A_0,g)}
.\end{equation}
The result of these operations depends on the remaining variables
$A_0$, $\delta A_{0,h}$, $g$, $\delta g$, and $\chi_h$.
This result equals the integrand
on the right hand side of the following formula
\begin{equation}
\w Z_k^\pert(A_0,\delta A_{0,h},g,\delta g)
 = \int_{\chi_h} e^{i<\chi_h,\delta A_{0,h} >}
    \w Z_k^{hl}(A_0,\delta A_{0,h},g,\delta g,\chi_h)
    \w Z_k^{sc}(A_0,\delta A_{0,h})
.\eqdef{smzk}\end{equation}
(Note that the exponential term in the integrand is just the
piece of $e^{i<\chi,\delta A>}$ which does not involve any of the
variables $\cA$, $\chi_d$, or $\delta_{vert} B$ already integrated over.)
$\w Z_k^{sc}$ is the semi-classical approximation obtained in doing
the operations above.  $\w Z_k^{hl}$ comes from the higher loop
Feynam rules.  It is possible to write it in the form
\begin{equation}
\w Z_k^{hl}(A_0,\delta A_{0,h},g,\delta g,\chi_h)
  = e^{ik {1\over 3!} <\part{\cJ},[\part{\cJ},\part{\cJ}]> }|_{\cJ=0}
    e^{{1\over 2ik} \Phi }
.\eqdef{fzkhl}\end{equation}
Here $\cJ\in [\w\Omega^*_M]_{(A_0,g)}$ and $\Phi$ is bilinear in
the pair $(\cJ,\chi_h)$.

\subsection{Translating Back to Language of Differential Forms}

When we translate back from the language of supermanifolds, we
obtain explicit formulas for
\begin{equation}
\w Z_k^{sc} \in \Omega^{|\cM|}(\cS)_{\mathrm{basic}}\otimes\PS(k)
                =\Omega^{|\cM|}(\cM)\otimes\PS(k)
                \subset\Omega^{|\cM|}(\w\cM)\otimes\PS(k)
\end{equation}
and
\begin{equation}
\w Z_k^{hl} \in
 \left[\mathop{\oplus}\limits_l
     \Omega^l(\w\cS;\Lambda^l(\w\Omega^1_h))_{\mathrm{basic}}
   \right] \otimes\IC[[{1\over k}]]
=\left[\mathop{\oplus}\limits_l
     \Omega^l(\w\cM;\Lambda^l(T\cM))
    \right] \otimes\IC[[{1\over k}]]
.\end{equation}The space $\Lambda^*(\w\Omega^1_h)$ which the forms are valued
in
just corresponds to the space of smooth functions of the supervariable
$$
 \chi_h\in\w\Omega^2_h=\left[\w\Omega^1_h\right]^*
$$

Equation \eqref{smzk}\ translates into the formula
\begin{equation}
\w Z_k^\pert
  = \w Z_k^{hl}\wedgeinner \w Z_k^{sc}
   \in \Omega^{|\cM|}(\w\cS)_{\mathrm{basic}}\otimes\PS(k)
        =\Omega^{|\cM|}(\w\cM)\otimes\PS(k)
.\end{equation}
The symbol $\wedgeinner$ indicates that the multi-vector valued form $\w
Z_k^{hl}$
is to act on the form $\w Z_k^{sc}$ by a combination of inner product with
multi-vectors and wedge products for forms.
$\w Z_k^{sc}$ is easily shown to be closed.  $\w Z_k^\pert$ can be
shown to be closed directly (i.e. without relying on the existence of
the integral \eqref{pathint}\ which is ill-defined in the case of quantum field
theory) by showing that $\w Z_k^{hl}$ is
closed under an extension of the exterior derivative
operator which acts on multi-vectors by taking exterior derivative
in the $\Met$ directions and divergence with respect to the volume
form $\w Z_k^{sc}$ on $\cM$ in the $\cM$ directions so that
\begin{equation}
      d_{\w\cM}(\w Z_k^{hl}\wedgeinner \w Z_k^{sc})
        = (d_{\w\cM}\,\w Z_k^{hl} )\wedgeinner \w Z_k^{sc}
.\eqdef{zkcl}\end{equation}
That $\w Z_k^{hl}$ is closed can be proved directly from the
prescription \eqref{fzkhl}.

This ends our overview of the formulation of
the finite dimensional perturbation series and the proof of its metric
independence in a way that does not involve any integrals that would become
infinite dimensional in the field theory problem.
Although we have been rather sketchy in this section and may have
used some perhaps unfamiliar machinery, it is hoped that the reader
can get a good feel for what's going on here by comparing with the
trivial example of the proceeding section.

\section{Application to Chern--Simons Theory}

In the case of Chern--Simons theory, \eqref{fzkhl}\ may be translated
to a definition of the following form, where the meaning of some of the symbols
is explained below\footnote{
  For simplicity, and because it does not effect whether or not
  $Z_k^{pert}$ is an invariant, we have not included the normalization
  factors and the shift in $k$ discussed in \cite{ASI}\ needed for
  a conjectured agreement with the asymptotic expansion of Witten's exact
  solution.
  }:
\begin{equation}
\w Z_k^{hl} = e^{c(k)\op{CS}_{grav}(g,s)}
  \int_{e^M}\Tr_k(e^{{1\over 2ik}\Phi } )
 \in\Omega^*(\w\cM;\Lambda^*(T\cM))\otimes\IC[[{1\over k}]]
.\eqdef{garble}\end{equation}
The term in front of the integral sign is the ``counterterm''.
In it, $\op{CS}_{grav}(g,s)$ is the Chern--Simons invariant of the metric
connection defined relative to the canonical (bi-)framing.
The quantity $c(k)$ is an expression depending only on $k$ and the structure
group $G$ of the principle bundle $P$.  For a direct
translation from the finite dimensional case, the counterterm wouldn't appear,
i.e. $c(k)$ would be $0$.  We will see shortly the the counterterm is
the same as was found necessary in the case of perturbation theory about
an acyclic flat connection \cite{ASI,ASII}\ and arises for the same reason.

The expression $e^M$ in \eqref{garble}\ stands for the union
\begin{equation}
e^M = \cup_{V=0}^\infty M[V]/S_V
,\end{equation}
where $M[V]$ is the compactification appearing in \cite{ASII} of the
configuration space of $V$ distincts points in $M$ and $S_V$ is the
action of the symmetric group on this space.  Using some elliptic operator
techniques and Hodge theory,  $\Phi$ is defined in terms of the natural
structures on the deformation complex discussed in the previous section.  The
restriction of $\Phi$ to $M[V]$ is a form\footnote{
    We use the notation in \eqref{phivsp}\ for simplicity.
    Strictly speaking $\Phi_V$ belongs to the $\cG$-invariant subspace of
    $\Omega^*(\w\cS;\Lambda^*(\w\Omega^1_h))\otimes
     \Omega^*(M[V]; A_V^*)^{S_V}$.
    }
\begin{equation}
\Phi_V \in \Omega^*(\w\cM;\Lambda^*(T\cM))\otimes
   \Omega^*(M[V]; A_V^*)^{S_V}
\eqdef{phivsp}\end{equation}
where
\begin{equation}
 A_V^*=\Lambda^*(\adp_1\oplus ...\oplus\adp_V)
\end{equation}
 with
 $\adp_i$ being the adjoint bundle to $P$ pulled back to $M[V]$
by the projection map from $M[V]$ onto the $i$'th copy of $M$.
$\Tr_k$ is a map (normalized by a power of $k$) from $A_V^*$ to $\IR$,
so that it makes sense to perform the integral in \eqref{garble}.
It is convenient to define
\begin{equation}
\Omega^*(e^M) = \bigoplus_V
  \Omega^*(M[V];A_V^*)^{S_V}
.\end{equation}
Note that the first exponential in \eqref{fzkhl}\ written in language
appropriate to Chern-Simons theory becomes the formal expression
\begin{eqnarray*}
  e^{{ik\over 3!} \int_{x\in M}
      \Tr(\part{\cJ(x)}\wedge[\part{\cJ(x)},\part{\cJ(x)}])}
\end{eqnarray*}
where $\cJ\in\Omega^*(M;\adp)$ and $-\Tr$ is an inner product on
$\Lie(G)$.  It is not suprising that this can be given a sensible
expression of the form \eqref{garble}.

In the case of perturbation theory about an acyclic connection considered in
\cite{ASII}\ (so that $\cM$ is a point), each of the
forms $\Phi_V$ is closed.  This allowed us to apply Stokes theorem
to calculate
\begin{equation}
d_\Met \left(\int_{M[V]} \Phi_V \right)
 = \int_{\del M[V]} (d_\Met + d_{M[V]}) \Phi_V
  -\int_{\del M[V]} \Phi_V\  =\  -\int_{\del M[V]} \Phi_V
.\eqdef{simpled}\end{equation}
The right hand side of this equation is called an anomaly
because it would vanish in the finite dimensional case.
Using an explicition description of $\del M[V]$ and
$\Phi_V|_{\del M[V]}$ to calculate the anomaly, we
found that $\w Z_k^{hl}$ could be made closed (i.e. metric independent) by
defining $c(k)$ appropriately.

When $\cM$ consists of more than a point,
the situation  turns out to be quite a bit more
complicated.  Even in the trivial example of \S 2,
 $\w Z_k^{hl}$ is only closed because of a cancellation among diagrams with
different numbers of vertices.  Since the only reason that the
finite dimensional cancellation might not apply in the field
theory case is due to the singularities that arise in the definition
of $\Phi$ and show up near points in $\del e^M$; we expect there
should be some way to reduce the anomaly calculation to some kind of
bondary integral.
In fact, with a bit of work, it turns out that
we may apply a sort of ``Stoke's theorem on $e^M$'' so that the anomaly is
reduced to a calculation on $\del e^M$ as was true in the
acylic case.

To formulate this ``Stoke's theorem on $e^M$'',
 we find an algebra $\cA^*$ (generated by a certain set of labelled graphs)
with an operator $\u D:\cA^*\to\cA^{*+1}$ and an algebra homomorphism
\begin{equation}
  I:\cA^*\rightarrow
  \Omega^*(\w\cM;\Lambda^*(T\cM))\otimes \Omega^*(e^M)
\end{equation}
This algebra homomorphism embodies the Feynman rules and their variation under
exterior derivative.
$\cA^V$ is generated by graphs which have $V$ vertices of a type that is
considered internal and is mapped by $I$ to
\begin{equation}
  \Omega^*(\w\cM;\Lambda^*(T\cM))\otimes\Omega^*(M[V]; A_V^*)^{S_V}
.\end{equation}
The operator $\u D$ is a sum
\begin{equation}
  \u D = \u D_0 + \u D_1
\end{equation}
 where $\u D_i$ increases the number of vertices by $i$.
Finally, we have
\begin{equation}
\int_{e^M} \Tr_k(I(\u D\, \u\omega))
  = \int_{e^M} \Tr_k( d_{\w\cM\times e^M} I(\u\omega) )
\qquad\hbox{for $\u\omega\in\cA^*$}
,\eqdef{dcomm}\end{equation}
where $d_{\w\cM\times e^M}$ is the exterior derivative operator
on $\w\cM\times e^M$
(acting on sections of $\Lambda^*(T\cM)$ by taking divergence
with respect to the volume form $\w Z_k^{sc}$,
see the comment above \eqref{zkcl}).

To give a rudimentary idea of the meaning of $\cA^*$, $I$, and $\u D$
let us begin by going back to the trivial example of \S 2.
Recall that the Feynman rules in \S 2 associated to the edge of a graph a
factor of $-H(x_0)^{-1}$ (we set $ik=1$ here for convenience).  The exterior
derivate on $\w\cS$ of this term is
\begin{equation}
d_{\w\cS} (-H(x_0)^{-1}) = (- H(x_0)^{-1})  [ H'(x_0) dx_0 ] (-H'(x_0)^{-1})
.\eqdef{dho}\end{equation}
Graphically, this can be represented as follows:
\begin{equation}
  {\u D}(\  \epsfbox{prop.eps}\  )\quad =\quad
    \raisebox{-25pt}{\epsfbox{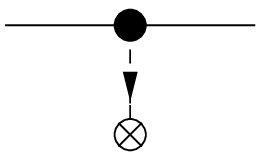} }
\eqdef{deredge}.\end{equation}
The operator $\u D$ acting on graphs
correspond to the exterior derivative operator $d_{\w\cS}$.
In the graph on the right in \eqref{deredge}, the top vertex is considered
internal and the bottom vertex is considered external.
The homomorphism $I$, capturing the Feynman rules,
associates a factor of $-H(x_0)^{-1}$ with the
solid edges as usual.  The dashed edge (with an orientation pointing down
indicated by the arrow) is given a factor of $1$.
The Feynman rule for the solid vertex is a factor of $H'(x_0)$, and the
Feynman rule for the bottom vertex is a factor of $dx_0$.
Thus we see that graphically the exterior derivative of an edge, which has
no vertices, is given by the labelled graph on the right in
\eqref{deredge}\ which has one internal vertex.
One can then proceed to find an algebra which is generated as a vector space
by graphs with various kinds of edges and vertices which contains the
orginal Feynman rules and has an operator $\u D$ capturing exterior
derivative.  In fact, there is more than one way to do this.  For example,
rather than introduce the graph on the right in \eqref{deredge}\ and the
Feynman rules above, we could simply have written the right hand side as
a single dashed edge with Feynman rule given by the right hand side of
\eqref{dho}.   However such a definition would fail to capture the fact that
the ingredients in the right hand side of \eqref{dho}\ appear in other
Feynman rules.  Such relations are essential in proving that the exterior
deriviatives of various Feynman diagrams (such as the ones in \eqref{itwo})
cancel.

Getting back to Chern--Simons theory,
what is remarkable is that we can find $\cA^*$, $\u D$, and $I$
satisfying \eqref{dcomm}\ and such that there exists
an element $\u\Phi\in\cA^*$ which is taken by $I$ to $\Phi$ and satisfies
the condition that $\exp({{1\over 2ik}\u\Phi})$ is
annihilated by $\u D$.
Applying \eqref{dcomm}\ to
$\u\omega = \exp({{1\over 2ik}\u\Phi})$ which is annihilated by $\u D$, we
obtain
\begin{equation}
0 = \int_{e^M} \Tr_k( I(\u D\, e^{{1\over 2ik}\u\Phi} ))
  = \int_{e^M} \Tr_k( d_{\w\cM\times e^M}\  e^{{1\over 2ik}\Phi } )
  = d_{\w\cM}\,\w Z_k^{hl}
    +\int_{\del e^M} \Tr_k( e^{{1\over 2ik}\Phi } )
.\end{equation}
We are thus left with a calculation of anomalies by
a boundary integral as in the acyclic case.  Although the definition
of $\Phi$ is quite a bit more complicated than in the acyclic case,
a simple calculation shows that the part of
$\Phi|_{\del e^M}$ which contributes to the anomaly calculation
is the same as in the acyclic case.  Thus the anomaly is the
same as in the acyclic case, and $\w Z_k^{hl}$ is indeed closed,
with the same counterterm as in the acyclic case.  Hence
$\w Z_k^\pert$ is closed and the perturbation series $Z_k^\pert$
is metric independent and therefore a diffeomorphism invariant.

\section{Concluding Remarks}

The result we have described here for Chern--Simons theory is a
paradigm that we expect should generalize to other quantum field
 theories.  Although complete details will appear elsewhere \cite{AXE},
we outlined how to formulate
perturbation theory about a component of the moduli space of instantons
(solutions to the equations of motion moduli gauge
invariance) on an arbitrary compact spacetime manifold  and
how to rigorously determine the anomalies to all orders.
The details of the calculation verify that the anomaly is in fact
of a universal nature, independent of the geometry of the instanton
moduli space.  This is similar to a result obtained by
Friedan \cite{FRIEDAN}
in studying sigma model perturbation theory and might be predicted
based on naive physical arguments.  A general formulation and proof
of this statement would be extremely useful.

The meaning of the form
$\w Z_k^{pert}\in\Omega^*(\cM\times\Met)\otimes\PS(k)$ may be
illuminated further by considering the case when our three manifold
$M$ equals a product $\Sigma\times S^1$ and
$\cM$ is a smooth component of the moduli space of flat
connections with structure group $G$ on
a Riemann surface $\Sigma$, identified with a smooth
component of the moduli space of flat connection on
$M$.  In this case, the path integral for the Chern--Simons theory
formally gives and Witten's exact solution \cite{WITTENI}\ actually
does give the result that $Z_k$ equals the dimension of the
Hilbert space obtained by quantizing the moduli space of flat $G$-connections
on $\Sigma$.  The contribution of $\cM$ to $Z_k$ then equals the dimension of
the space of holomorphic sections
of $\cL^{\otimes k}$, where $\cL$ is a certain holomorphic line
bundle on $\cM$.
(Recall that $\cM$ receives a K\"ahler structure once a
metric on $\Sigma$ is chosen and that  $\cL$ is a line bundle whose curvature
$2$-form equals the K\"ahler form of $\cM$.)
Using the Riemann-Roch index theorem and a vanishing
theorem, we find the exact answer (for large enough $k$) is a
obtained by integrating  the index density
 over $\cM$.  Assuming the general
conjecture that the invariants we have defined here do indeed give
the contributions of $\cM$ to the large $k$ asymptotics of the exact
solution, we find
\begin{equation}
  \int_\cM\op{Td}(T^{(1,0)}\cM)\wedge\op{ch}(\cL^k)
          = \int_\cM \w Z_k^\pert
\end{equation}
Note that the we expect the asymptotic series on the right to converge
to the function on the left since the latter is just a polynomial in $k$.
It is natural to conjecture that the restriction of $\w Z_k^\pert$
to $\cM\times\Met(\Sigma)$, with an appropriate embedding of $\Met(\Sigma)$ in
$\Met(M)$, equals the index density above.
In other words, this conjecture  says that $\w Z_k^\pert$ is a
manifestly three-dimensionally invariant version of
 a quantity arising from a two-dimensional index theorem.

\appendix
\section{Recap of the Feynman Rules}

We now briefly recall the Feynman rules for calculating the
stationary phase approximation near the origin for an integral
of the form
\begin{equation}
  Z_k = \int_{A\in\IR^n} O(A) e^{i k S(A)}
\end{equation}
where $S$ and $O$ are smooth function on $\IR^n$, $S$ has a
non-degenerate critical point at the origin, and $O$ has
compact support (so that the integral converges).  We will not
give a proof of this formulation here, although it can be
derived from the description of the stationary phase approximation
given in \cite{ARNOLD}, for example.  We will simply recall the
formal derivation due to Feynman which can be found in many textbooks
on quantum field theory, for example \cite{RAMOND}.

We begin by writing down the Taylor expansion for $S$ about the origin,
\begin{eqnarray}
 S(A) &=&  S(0) + \half <A, H A> + V(A)  \\
 V(A) &=& \sum_{v=3}^\infty {1\over v!}
      {\del^v S(0)\over\del A^{i_1}...\del A^{i_v}} A^{i_1} ... A^{i_v}
\end{eqnarray}
where $H$ is the Hessian of $S$ at the origin, considered as a linear
map from $\IR^n$ to itself (or more precisely its dual space).
Next, we introduce a variable $J$ in $\IR^n$ (or again, more precisely its
dual space) and interpret expressions such as $O(\part{J})$ as a formal
power series obtained by plugging into the Taylor series for $O$,
\begin{equation}
O(\part{J}) = \sum_{v=0}^\infty {1\over v!}
     {\del^v O(0)\over\del A^{i_1}...\del A^{i_v}}
           \part{J^{i_1}} ... \part{J^{i_v}}
\end{equation}
It is apparent formally and in fact true that we may write
\begin{eqnarray}
Z_k        &=& e^{ik S(0)}\int O(A) e^{i k V(A)} e^{ik <A, H A>} \\
Z_k^\pert  &=& e^{ik S(0)} \left[ O(\part{J}) e^{ik V(\part{J})}\right]_{J=0}
       \int e^{ik <A,H A> + <J, A>}
.\end{eqnarray}
The final integral is defined by adding a small imaginary part to $H$ so that
the integral converges and then taking the limit as the imaginary part
goes to zero.  The result may be calculated by completing the square and
recalling the value of a Gaussian integral.  This yields
\begin{eqnarray}
  Z_k^\pert &=& Z_k^{hl}\, Z_k^{sc}
\\
  Z_k^{sc}  &=& \left({2\pi\over k}\right)^{n/2}
                {{e^{\pi i\sign(H)\over 4}} \over |\det(H)|} e^{ik S(0)}
\\
  Z_k^{hl}  &=& \left[ O(\part{J}) e^{ik V(\part{J})}\right]_{J=0}
                e^{-{1\over ik}\half <J, H^{-1} J >}
\\
            &=& \sum_{I=0}^\infty\sum_{V=0}^\infty
                   \left[O(\part{J}) {\left[ ik V(\part{J})\right]^V\over V!}
                   \right]_{J=0}
                    {\left[ -{1\over ik}\half <J,H^{-1} J>\right]^I\over I!}
.\end{eqnarray}

With a little thought one can see that the term inside the final sum
can be expressed in the form
$\sum_\Gamma {I(\Gamma)\over S(\Gamma)}$, where the sum runs over
all graphs with $I$ edges, $V$ unmarked vertices all of which have valency
at least three, and $1$ marked vertex.  The factor $S(\Gamma)$ is the order
of the automorphism group of $\Gamma$.
The ``Feynman amplitude'' $I(\Gamma)$ is obtained by the following rules.
First, one
labels each end of each edge by the name of an index running from $1$ to $n$.
Then one writes down the product of the following factors:
a factor of $-{1\over ik} (H^{-1})^{jk}$ associated with each edge, where
$j$ and $k$ are the index names labelling the ends of the edge;
a factor of $ik {\del^v S(0)\over\del A^{i_1}...\del A^{i_v}}$ associated to
each unmarked vertex, where $v$ is the valency of the vertex and
$i_1$,...,$i_v$ are the index names labelling the edge ends incident on
the vertex; and a factor of $\del^v O(0)\over\del A^{i_1}...\del A^{i_v}$
associated to the marked vertex, where $v$ is its valency and $i_1$,...$i_v$
are the incident edge end labels.  Finally, one obtains the
``Feynman amplitude'' $I(\Gamma)$ by summing the product of all these factors
as each of the indices runs from $1$ to $n$.
The empty graph has $I(\Gamma)=1$ and $S(\Gamma)=1$.  All other graphs
contributing to $Z_k^{hl}$ have more than $1$ loop

For example, here are some pictures and
their associated factors.  The third picture, for example, represents
a marked vertex of valency one with incident edge end label $j$.
\begin{eqnarray}
  i\ \epsfbox{prop.eps} \ j
  && -\frac{1}{i k} (H^{-1})^{ij}
  \\
  \lower .25in\hbox{\epsfbox{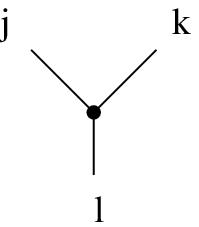}}
  && i k {\del^3 S(0)\over\del A^j\del A^k\del A^l}
  \\
  \epsfbox{stickball.eps}\ j
  && {\del O(0)\over\del A^j}
\end{eqnarray}

As a final example, the graph
\begin{equation}
  \epsfbox{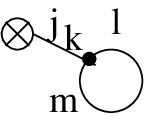}
\end{equation}
has Feynman amplitude
\begin{equation}
\sum_{j,k,l,m}
   \left[-\frac{1}{i k} (H^{-1})^{jk}\right]
     \left[-\frac{1}{i k} (H^{-1})^{lm}\right]
     \left[{\del O(0)\over\del A^j}\right]
     \left[ i k {\del^3 S(0)\over\del A^j\del A^k\del A^l}\right]
\end{equation}

Finally, we remark that when $O$ equals the constant function $1$, then
we may just as well delete the marked vertex and sum
over ordinary unmarked graphs.  In the trivial example of \S 2, we refer to
the unmarked vertices as ``internal vertices'' and to the marked vertex
as an ``external vertex''.  In that example, the indices are not even
necessary since $n=1$.

\newpage
\def\article#1#2#3#4#5{{\it #1}, #2 {\bf #3} (#4) #5}


\newpage
\section*{List of Notations}
In the list below, the general description is for the finite dimensional
context. Expression in square brackets apply in the
Chern-Simons theory context.  Expression in angle brackets apply to
a general Euclidean quantum field theory.

\bigskip

\noindent
\begin{tabular}{llp{5in}}
$M$      & [closed, oriented $3$-manifold] \\
$G$      & [compact Lie group] \\
$P\to M$ & [principal $G$ bundle] \\
$\cF$    & manifold integrated over \ft{space of fields} [connections on $P$]
\\
$S$      & Morse-Bott function on $\cF$ \ft{action} [Chern-Simons invariant]\\
$k$      & real parameter\ft{inverse Planck constant} [postive integer level]\\
$\cG$    & symmetry group acting on $\cF$ [gauge transformations of $P$] \\
$\mu$    & volume form on $\cF$ \ft{formal path integral measure} \\
$\cS$    & set of critical points of $S$ \ft{instantons}
           [flat connections on P] \\
$\cM$    & $\cS/\cG$ \ft{instanton moduli space} \\
$\cH$    & subgroup of $\cG$ to which all isotropy subgroup are conjugate \\
$A_0$    & point in $\cS$ [flat connection on $P$] \\
$[A_0]$  & point in $\cM$ [gauge equivalence class of $A_0$] \\
$\Met$   & space of $\cG$-invariant metrics on $\cF$
	      [Riemannian metrics on $M$] \\
$g$      & element of $\Met$ \\
$N\to\cS$& normal bundle to $\cS$ in $\cF$ \\
$\w\cS$  & $\cS\times\Met$ \\
$\w N$      & $N\times\Met$ \\
$E:\w N\to\cF$
         & evaluation map (exponential map with variable $g$)\\
$\int^{pert}$	& perturbation series (stationary phase approx.)
		  for integral \\
$\PS(k)$    	& ring of perturbation series \\
$I(\Gamma)$     & Feynman amplitude associated to a graph $\Gamma$ \\
$S(\Gamma)$     & symmetry factor of a graph $\Gamma$ \\
$Z_k$    	& basic integral being considered \ft{partition function}\\
$Z_k^{pert}$	& stationary phase approximation to $Z_k$
		  [invariant we define] \\
                &  \ft{perturbation series to be regularized} \\
$\w Z_k^{pert}$ & form on $\w\cM$; integrated over $\cM$ yields $Z_k^{pert}$\\
$\w Z_k^{sc}$   & semi-classical part of $\w Z_k$ \\
$\w Z_k^{hl}$   & higher-loop contribution to $\w Z_k$\\
$\w I_l$        & $l$-loop piece of $\w Z_k^{hl}$\\
\end{tabular}

\begin{tabular}{llp{5in}}
$\wedgeinner$ & combination of wedge product and interior product operators \\
$\Omega^*_M$ & deformation complex for $\cM$ [$\Omega^*(M,\adp)$] \\
$D^*_{A_0}$ & differential on $\Omega^*_M$ at $A_0\in\cS$
              [twisted exterior derivative] \\
$T_{A_0}$   & infinitesimal action of $\cG$ at $A_0$ (equals $-D^0_{A_0}$)\\
$H(A_0)$    & Hessian of $S$ at $A_0$ (equals $D^1_{A_0}$) \\
$\w\Omega^*_M$ & bundle over $\w\cS$ of deformation complexes \\
$\w\Omega^*_h, \w\Omega^*_d, \w\Omega^*_\delta$
	    & harmonic, exact, and coexact subbundles of $\w\Omega^*_M$ \\
$\cA^*$	& [algebra generated by labelled graphs] \\
$I$	& [Feynman rule homomorphism from $\cA^*$ to differential forms] \\
$\u D$  & [analogue of exterior derivative acting on $\cA^*$] \\
$M[V]$  & [compactification of $M^V\setminus\{\mbox{all diagonals}\}$] \\
$e^M$   & [$\cup_{V=0}^\infty M[V]/S^V$, a closure of the set of finite
subsets of $M$] \\
\end{tabular}



\begin{thebibliography}{99}

\bibitem{ARNOLD}{
 V.I.\ Arnold, S.M.\ Gusein-Zade, A.N.\ Varchenko,
 {\it Singularities of Differential Maps, Volume II},
 Birkhauser, Boston, 1988.
}

\bibitem{AXE}{S.\ Axelrod, to appear.}

\bibitem{ASI}{
  S.\ Axelrod and I.\ M.\ Singer,
 {\it Chern--Simons Perturbation Theory}, Proc. XXth DGM Conference
 (New York, 1991) (S. Catto and A. Rocha, eds) World Scientific,
 1992, 3--45.
}

\bibitem{ASII}{
 S.\ Axelrod and I.\ M.\ Singer,
 \article{Chern--Simons Perturbation Theory, II}
 {J.~Diff.~Geom}{39}{1994}{173-213}.
}

\bibitem{DEWITT}{
 B.S. DeWitt,
 {\it Supermanifolds}, Cambridge University Press, Cambridge, 1984.
}

\bibitem{FRIEDAN}{
  D. Friedan,
  \it{NonLinear Sigma Models in $2+\epsilon$ dimensions},
  Annals of Physics, Vol. 163, No. 2, Sept. 1985
}

\bibitem{RAMOND}{
  P. Ramond,
  {\it A Primer on Quantum Field Theory}, Addison-Wesley, New York, 1990
}

\bibitem{WITTENI}{
 E.\ Witten,
 \article{Quantum field theory and the Jones polynomial}
   {Comm.  Math. Phys.}{121}{1989}{351-399}.
}

\end{thebibliography}
\end{document}